\documentclass[fleqn,pre,twocolumn,aps]{revtex4-1}
\usepackage{float,graphicx,amsmath,amssymb,subfigure,hyperref,color,textcomp}

\newcommand{\del}{\partial}
\newcommand{\m}[1]{\boldsymbol{\mathsf{#1}}}
\renewcommand{\v}[1]{\mathbf{#1}}
\renewcommand{\d}[1]{\,{\rm d}#1}
\newcommand{\D}[1]{{\rm D}#1}

\newcommand{\strain}{\dot{\boldsymbol \gamma}}
\newcommand{\strainnorm}{|\strain|}

\begin{document}

\title{Head-on collisions of dense granular jets}
\author{Jake Ellowitz}
\email{ellowitz@uchicago.edu}
\affiliation{The James Franck Institute and the Department of Physics, University of Chicago, Chicago, IL 60637}
\date{\today}

\begin{abstract}
When a dense stream of dry, non-cohesive grains hits a fixed target, a collimated sheet is ejected from the impact region, very similar to what happens for a stream of water. In this study, as a continuation of the investigation why such remarkably different incident fluids produce such similar ejecta, we use discrete particle simulations to collide two unequal-width granular jets head-on in two dimensions. In addition to the familiar coherent ejecta, we observe that the impact produces a far less familiar quasi-steady-state corresponding to a uniformly translating free surface and flow field. Upon repeating such impacts with multiple continuum fluid simulations, we show that this translational speed is controlled only by the total energy dissipation rate to the power $1.5$, and is independent of the details of the jet composition. Our findings, together with those from impacts against fixed targets, challenge the principle of scattering in which material composition is inferred from observing the ejecta produced during impact.
\end{abstract}
\maketitle

\begin{figure*}[ht]
\centering
\includegraphics[width=\textwidth]{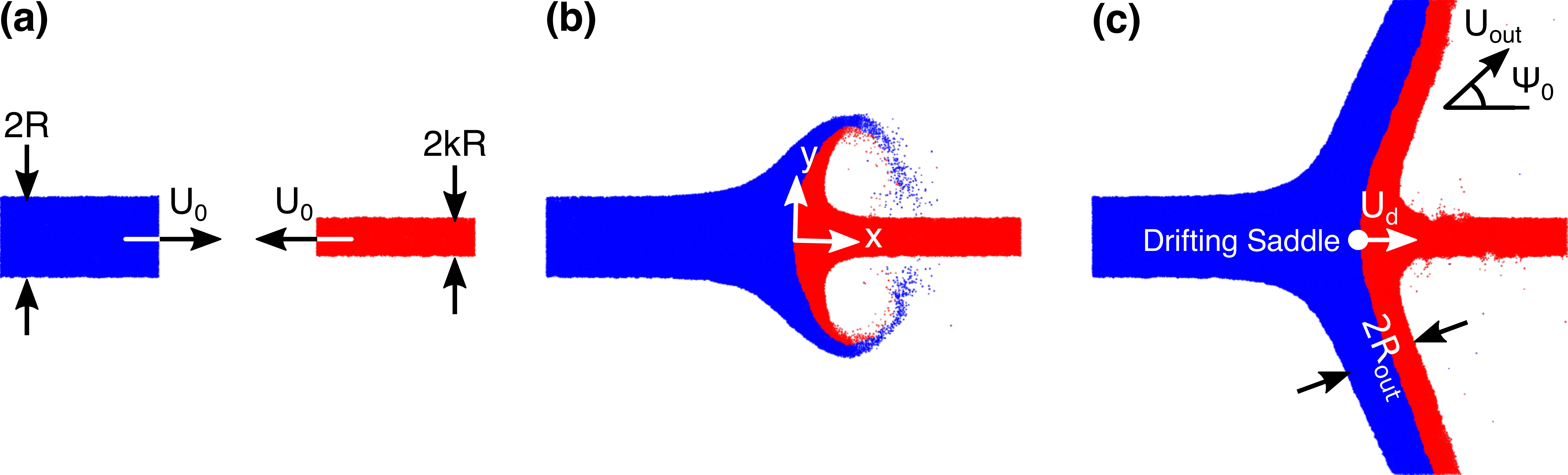}
\caption{\label{fig:schematic}
(Color online) Two-dimensional head-on impact of two jets produces a steady-state flow with collimated ejecta and a steadily drifting flow configuration. {\bf (a)} In two dimensions, A jet of width $2R$, shown in blue, is incident against a jet of with $2kR$, shown in red. Both jets are initially at $81\%$ packing fraction and traveling toward each other at the same speed $U_0$. Here, $R = 100R_{\rm G}$, where $R_{\rm G}$ is the grain width, and $k = 0.465$. {\bf (b)} Snapshot at $t = 2 R/U_0$ after the two jets directly collide. Here we indicate the coordinates $x$ and $y$ whose origin is located at the point of initial contact. {\bf (c)} Snapshot at $t = 10 R/U_0$, where the central saddle point, shown with a dot, is drifting toward the smaller oncoming jet with drift speed $U_d$. Here, $U_d = 0.052 U_{\rm 0}$. Additionally, the impact produces 2 collimated ejecta streams with widths $2R_{\rm out}$ and ejecta velocities with magnitude $U_{\rm out}$ exiting at an angle $\Psi_0$ from the horizontal. Note $\Psi_0$ is not quite the same as the spatial angle formed between the ejecta jet and the horizontal due to the fact that the ejecta streams are drifting alongside the saddle point.
}
\end{figure*}

\section{Introduction}

If we collide two ideal fluid streams head-on in two dimensions, one larger than the other but both the same speed, we would observe a steady state with two equally sized and inclined streams leaving the impact region. This is because the larger jet has more incident momentum than the smaller, and these ejecta are how the system accommodates for the momentum excess. This steady-state is achieved because ideal fluids are time-reversible and is thus the reverse of two equal-width jets incident at an angle. Due to symmetry, this reversed case is in turn equivalent to a jet impinging onto an infinite wall, a situation that surely reaches a steady-state. Hence, head-on collisions of ideal-fluids reach a steady-state where the ejecta compensate for the nonzero incident momentum.

Since this stationary steady-state impact owes its existence to the fact that the flow field is entirely reversible, breaking reversibility by including dissipation, no matter how small, should destroy the stationary solution. Intuition would then suggest that the head-on impact of dissipative jets should thus produce collective motion in the central impact region. Numerical simulations presented in the current study verify this expectation by demonstrating that a wide variety of dissipational jet impacts produce quasi-steady-state flows whose entire flow field and free surface translate at a constant drift speed. This drift speed is shown to increase as power law relative to the dissipation with exponent $1.5$. More surprisingly, we find that fluids with widely varying rheologies all collapse onto the same power law: the behavior of this bulk motion depends only on the amount of dissipation in the impact, and is independent of the dissipation mechanism or the constitutive material of the incident jets.

Our findings are at the intersection of two previously studied but disparate jet impact systems: that of directly incident turbulent Newtonian streams and that of jet impact against fixed targets. In the case of the impact of turbulent jets, previous experiments identified that in fixed-length cavities, the impact stagnation point shifts toward the nozzle with lower incident momentum flux, whose offset depends on the incident momentum fluxes and nozzle separation~\cite{zhi-gang_offset, li_stagnation_offset, li_axisymmetric_offset, li_turbulence_factor}. These studies were focused on the stagnation point shifts, and reach steady state flows in the lab frame due to the fixed nozzle separation. Our study is different because we investigate the offset mechanism and transient directly and are able to completely remove the nozzle effects.

Regarding impacts against fixed targets, previous workers showed granular and high speed liquid jets astoundingly both produce quite similar coherent ejecta sheets~\cite{clanet_water_bell, clanet_water_bell_review, xiang_original}. Because the flow from a high speed liquid impinging a fixed target is similar to a perfect fluid, this led to a discussion of whether or not a granular jet impact against a target behaves like a perfect fluid~\cite{muller_granular, xiang_original, sano_hayakawa_first, sano_hayakawa_second, oblique}.
Despite the similar ejecta from the target impacts, it was later shown that granular jet impact against a target produces a large, stagnant dead-zone, and that furthermore inserting large dead-zone-like structures into perfect fluids barely affected the ejecta~\cite{still_water}.
This dead zone, which is reminiscent of those found in dilute flows~\cite{swinney_shock, amarouchene_dilute_deadzone}, is a stark contrast to the stagnation point in perfect fluid impacts, yet the ejecta remained generic in its presence, showing that the jet composition and internal structure have little effect on the ejecta when the impact is cold and incompressible. 

Directly colliding fluid streams furthermore parallel multiple applications in manufacturing processes. Opposed jet micro reactors have been utilized to facilitate combustion and to produce highly controlled chemical reactions for polymer processing, nanoparticles, and the careful synthesis of organic compounds in the pharmaceutical industry~\cite{korusoy, johnson_opposed_jet, li_stagnation_offset, li_axisymmetric_offset, kumar_mixing}. Furthermore, opposing jets are employed in drying particles and fluid absorption~\cite{berman_absorbing_opposing, mujumdar_drying} and mineral extraction~\cite{dehkordi_extraction}. The present findings are relevant to the mechanics by which the impact center shifts between nozzles in these diverse industrial examples, in particular highlighting the effects of dissipation which have not yet been considered deeply, and are explored in this paper.

\section{Methods}

In this paper, we consider the two-dimensional head-on collision of two dissipative jets of different sizes: one jet of width $2R$ and a smaller jet of width $2kR$, where $0\leq k \leq 1$. The two jets are incident such that their centers align. Their initial velocities are both $U_0$ in magnitude, and are opposite in direction so that the two jets collide directly (Fig.~\ref{fig:schematic} (a)).
To simulate these impacts we used two separate numerical approaches: the first being a discrete particle simulation of dense granular jets~\cite{guttenberg_method, guttenberg_dissipation, still_water, oblique}, and the second being the continuum dynamics of colliding jets using the multi-phase, incompressible fluid flow solver {\sc gerris}~\cite{gerris_original, gerris_surface_tension, lagree_collapse, gerris}.

The discrete particle simulations consist of simulating the motion and collisions of hundreds of thousands of rigid spheres with specified coefficients of restitution and friction. We conduct our discrete particle simulations using rigid grains, in line with experiments which used copper and glass beads with negligible deformation during impact~\cite{xiang_original, still_water}. The discrete particle simulations utilize a hybrid timestep- and event-driven method designed to handle dense granular flow efficiently and accurately~\cite{guttenberg_method, guttenberg_dissipation}. This method has been used in many recent studies of granular jet impact, and has also been quantitatively validated against experimental granular jet impact flow configurations~\cite{still_water}.

We prepare our discrete particle jets at $81\%$ packing fraction, and because our simulations are in two dimensions, we use polydisperse grains to prevent spurious crystallization. The grains we use have radii uniformly distributed from $0.8R_{\rm G}$ to $1.2 R_{\rm G}$, where $R_{\rm G}$ is the average grain size. In this paper, we use a grain size $R_{\rm G} = R/100$. The jet-width-ratio $k$, restitution and friction coefficients are varied and the effects on the dynamics of head-on collision are explored.

The continuum simulations of colliding streams are conducted for liquid jet impacts, where the shear stress is proportional to the shear rate, for frictional fluid models of granular jets, where the shear stress is proportional to the pressure, and a shear thickening fluid whose effective viscosity increases with the shear rate. 

Continuum solutions in {\sc gerris} utilize an incompressible volume of fluid method to solve multi-phase flow while providing significant flexibility for specification of the deviatoric stress tensor~\cite{gerris_original, gerris_surface_tension, lagree_collapse}. Ideally, we are solving Cauchy's momentum equations with incompressibility
    \begin{equation}
    \rho \frac{\D \v u}{\D t} = \boldsymbol \nabla\cdot\boldsymbol \sigma,\quad \boldsymbol \nabla \cdot \v u = 0
    \end{equation}
where ${\rm D}/{\rm D}t = \del_t + \v u\cdot\boldsymbol\nabla$ is the advective derivative, $\rho$ is the jet density, $\v u$ is the fluid velocity field, and $\boldsymbol\sigma$ is the stress tensor. The initial conditions are
    \begin{equation}
    \v u = \left\{ \begin{array}{lr} U_0 \hat{\v x} & x < 0\\ -U_0 \hat{\v x} & x > 0\end{array} \right.
    \end{equation}
with an initial jet free surface corresponding to the two impinging jets right when they make first contact. The boundary condition on the free surface corresponds to zero-stress with $\boldsymbol\sigma\cdot\hat{\v n} = \v 0$, where $\hat{\v n}$ is the surface normal. The boundary conditions within the jet are $\v u = U_0\hat{\v x}$ for $x\rightarrow-\infty$ and $\v u = -U_0\hat{\rm x}$ for $x \rightarrow \infty$. 

Our stress tensor is in general given by
    \begin{equation}
    \boldsymbol \sigma  = -p \m 1 + \boldsymbol \tau
    \end{equation}
where $\m 1$ is the identity matrix, $p$ is the pressure and $\boldsymbol \tau$ is the deviatoric stress. In the cases presented in this paper, we consider three deviatoric stress tensors: that of Newtonian liquid flow, where $\boldsymbol\tau_{\rm Liq} = \eta \strain$, that of a frictional fluid, where $\boldsymbol\tau_{\rm FF} = \mu p \strain/\strainnorm$, and that of a shear thickening fluid $\boldsymbol\tau_{ST} = \kappa\strainnorm^{1/2}\strain$. Here, $\strain = \boldsymbol\nabla\v u + \boldsymbol\nabla \v u^{\rm T}$ is the rate of strain tensor, and $\strainnorm = \sqrt{\strain\boldsymbol :\strain/2}$.

For the Newtonian rheology, $\eta$ is the dynamic viscosity, for the frictional fluid, $\mu$ is the dynamic frictional coefficient, and for the shear thickening fluid $\kappa$ is the flow consistency index. Speaking in particular to the frictional fluid, this frictional fluid model has been shown to reproduce simulations and experiments of dense granular jet impact, and is therefore a natural choice for this study~\cite{still_water, oblique}. We additionally considered the Newtonian and shear thickening fluids because their stress tensors are significantly different from dry granular flow. These differences are key in our results showing that widely varying materials produce similar head-on impact responses.

Even though our idealized problem consists of free-surface jets of a single fluid type colliding in free-space, the volume of fluid method that {\sc gerris} utilizes to solve the equations based on the boundary and initial conditions above requires a surrounding fluid occupying the free space outside the jets in our idealized problem. In order to minimize the effects of the surrounding fluid, we set its density to $\rho_{\rm surrounding} = \rho/250$, and we set its deviatoric stress to be Newtonian with viscosity $\eta_{\rm surrounding} = 10^{-3} R U_0 \rho_{\rm Inter}$. We conducted our simulations with refinement 26 grid points across the jet width. 

We have extensively varied the grid refinement, the surrounding fluid viscosity, and surrounding fluid density to ensure that the above parameters are in a regime where the effects of the surrounding fluid and grid discretization are minimal. Specifically, using a frictional fluid with $\mu=0.3$ as a representative impact, we found that there was a variation in the drift speed of roughly $0.1\%$, and a variation of the energy dissipation rate of roughly $2\%$.

\begin{figure}[ht]
\centering
\includegraphics[width=\columnwidth]{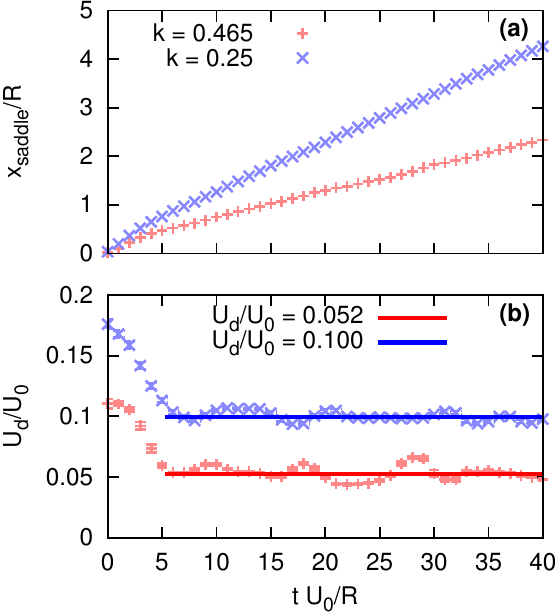}
\caption{\label{fig:udt}(Color online) Following a short transient, the central saddle point drifts at a relatively steady speed $U_d$. Normalized {\bf (a)} saddle position and {\bf (b)} corresponding drift speed versus time for collision between two jets at $k=0.465$ (red) and $k = 0.25$ (blue). The impact initially occurs at $x_{\rm saddle} = 0$. Following a short transient of duration $\sim 4 R/U_0$, the saddle point translates linearly in time. The direct measurement of the drift speed obtained by differentiating the saddle position. The drift speed starts high in the transient state and then fluctuates about the mean drift speed after steady drift has been reached. The drift speeds above are $U_d/U_0 = 0.052$ for $k = 0.465$ and $U_d/U_0 = 0.100$ for $k = 0.25$.
}
\end{figure}

\begin{figure*}[ht]
\centering
\includegraphics[width=\textwidth]{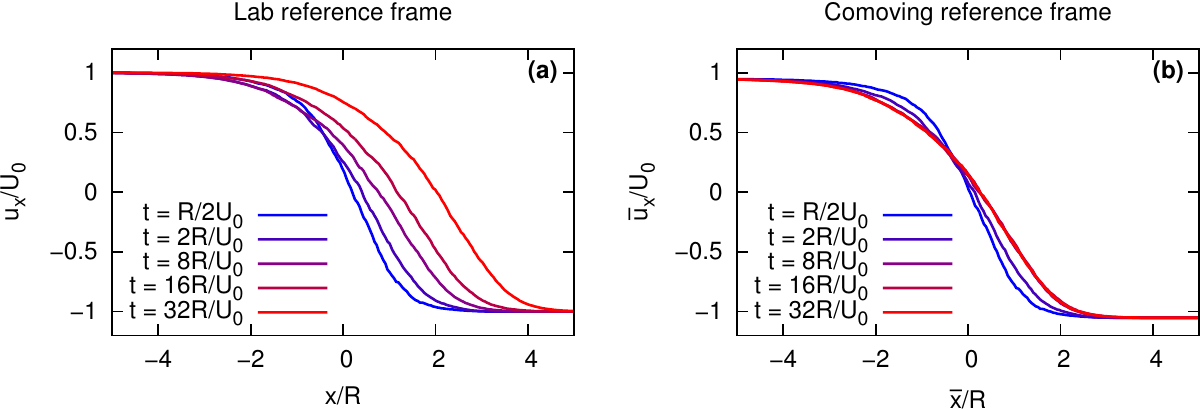}
\caption{\label{fig:collapse}
(Color online) The entire flow configuration arising from the head-on collision of unequal jets drifts steadily in time. {\bf (a)} As viewed in the fixed lab reference frame, we show horizontal velocity $u_x/U_0$ versus position $x/R$ along the line running through the centers of the jets defined by $y = 0$. Velocities are shown at a variety of times, ranging from very soon after impact (blue) to long after impact (red). The velocity fields at these different times does not coincide when viewed from the lab frame. {\bf (b)} Horizontal velocity profile $\bar x/R$ along the center line as viewed in the comoving reference frame which is translating steadily at the drift speed $U_d$ along with the drifting saddle. All of the profiles from (a) are included, where all of the later ($t > 2 R/U_0$) snapshots directly overlap in the comoving frame. Hence, excluding the profiles obtained in the initial transient, the flow configuration translates at a constant drift speed $U_d$, remaining preserved as it translates. 
}
\end{figure*}

\section{Results}

To briefly summarize our results, using discrete particle simulations, we observe the resulting flow from the collision of two unequal-width dense jets composed of dry, rigid grains in two dimensions. We reveal that after an initial transient, the impact produces a quasi-steady state corresponding to a uniformly translating flow field and free surface at a fixed drift speed. This steady drift speed increases as the size of the larger jet with respect to the smaller jet increases, and occurs alongside momentum being absorbed into the drifting impact center. We then consider multiple continuum jet impacts and find that the drift behavior was extremely robust, to the extent that the total energy dissipation controls the drift speed regardless of the significantly different dissipation mechanisms present in our continuum and discrete particle simulations. This dependence manifests itself via the drift speed robustly scaling versus the dissipation rate with a $1.5$ exponent. 


\subsection{Steady saddle drift}

To begin, we consider two unequal-width granular jets with $k=0.465$, and whose grains have coefficient of restitution $0.9$ and coefficient of friction $0.2$ (see Fig.~\ref{fig:schematic}(a)). Unless otherwise stated, these are the values we use in our representative simulation examples in our discrete particle jet impacts. In Fig.~\ref{fig:schematic} we show early snapshots (a) just before impact, (b) just after impact, and (c) after the impact flow configuration has been well established. The collision produces some familiar behavior, namely that two collimated ejecta streams are produced as in the case of impact against a fixed target~\cite{xiang_original, still_water, xiang_oblique, oblique}. The collimated ejecta are traveling with velocity $U_{\rm out}$ at an angle $\Psi_0$ relative to the horizontal. 

However, we also find notably unfamiliar behavior in this system: the steady collision of the two incident streams produces a steadily translating mode in which the impact center and ejecta streams drift at a constant speed $U_d$. As a result of this drift velocity, the spatial angle traced by the ejecta stream relative to the horizontal (Fig.~\ref{fig:schematic}(c)) is slightly different from $\Psi_0$, the ejecta velocity angle.
In contrast, the more studied case of impact against a fixed target, there is always a fixed steady state where the ejecta streamlines are equal to their pathlines. But, in general, this is not true for head-on collisions of two jets, since they obtain a quasi-steady-state where the flow field and free surface steadily translate.

Let us define the saddle position $x_{\rm saddle}$ as the central point in the impact at the interface where the grains of the large jet meet the grains of the small jet. In Fig.~\ref{fig:schematic}(c), this is where the blue and red jets interface along the line $y=0$ and moving toward the smaller incident jet. The saddle center is initially situated at $x = 0$ at the point where the two jets first make contact (Fig.~\ref{fig:schematic}(b)). Because the granular jets are dense, and therefore cold~\cite{still_water, oblique}, there is negligible diffusion and mixing between the jets and their interface is clearly defined. In Fig.~\ref{fig:udt} (a) we track the saddle position over time, and find that following a short transient, the two jets produce a steady central motion where the saddle position moves linearly in time. 

We instantaneously measure the drift speed over time by computing the time derivative of the saddle position by fitting a slope to 10 measurements of $x_{\rm saddle}$ over a time frame $R/U_0$. These drift speed measurements are shown in Fig.~\ref{fig:udt}(b). In Fig.~\ref{fig:udt}(a) ,we graph only one $x_{\rm saddle}$ for $10$ measurements conducted in from our simulations so that each $x_{\rm saddle}$ corresponds to one instantaneous $U_d/U_0$ in Fig.~\ref{fig:udt}(b). This ensures that each $U_d$ is independently measured from all other $U_d$ shown. In the discrete particle jets, we see that there are slight fluctuations in the rate of motion of the saddle point about the average drift speed on the order of $U_0/100$, with an average drift speed $U_d/U_0 = 0.052$ to the right when $k = 0.465$ and $U_d/U_0 = 0.100$ when $k = 0.25$, where the large jet is pushing the small jet. Thus, smaller $k$ produces a larger $U_d$.

As we next show, not only does the saddle drift steadily in time, but the entire flow field drifts along with the saddle steadily in time. We measure the flow field along the jet centers by spatially binning grains into square boxes of linear dimension $4.5 R_{\rm G}$ along the center of the jet at $y=0$ and averaging the particle speeds within each bin. A typical bin contains $5$ particles. 

In Fig.~\ref{fig:collapse}(a), we show the instantaneous horizontal velocity field $u_x$ along the center line defined at various points in time. We see the velocity in the large left jet ($x \lesssim -4 R$) is $U_0$, and the velocity at large $x$ in the smaller jet ($x \gtrsim 4R$) is $-U_0$. The velocity fields at different points in time as viewed from this fixed lab frame appear distinct from one another.

We now enter the comoving frame whose origin is moving to the right at the drift velocity $U_d$ obtained from tracking the saddle point motion as in Fig.~\ref{fig:udt}. Here we find a dramatically simpler picture than in the lab frame. Let us denote coordinates in the comoving frame with a bar above the variable, notation that will be continued through the rest of this paper. The comoving frame coordinates in this notation are
    \begin{align}
    &\bar x = x - U_d t \\& \bar y = y\\
    &\bar u_x = u_x - U_d \\& \bar u_y = U_y
    \end{align} 
In the comoving frame, neglecting the short transient of duration $\sim 4R/U_0$, the velocity fields are indistinguishable. The clear conclusion here is that when two unequal-width granular jets collide, they produce a quasi-steady-state in which the entire flow field uniformly drifts in time at a fixed drift speed $U_d$ in the lab frame. In the comoving frame, the impact produces a steady-state, where partial time derivatives of the velocity field or interface locations ($\del_t$) vanish, and the streamlines match the pathlines, resulting in the spatial angle traced by the ejecta relative to the horizontal to match the velocity ejecta angle.

Having established that a central drift emerges during impact of two head-on granular jets of unequal widths, we next explore the effects of jet width, and will profile the flow configuration as $k$ is varied.

\subsection{Characterization of post-impact flow configurations}

\begin{figure}[t]
\centering
\includegraphics[width=\columnwidth]{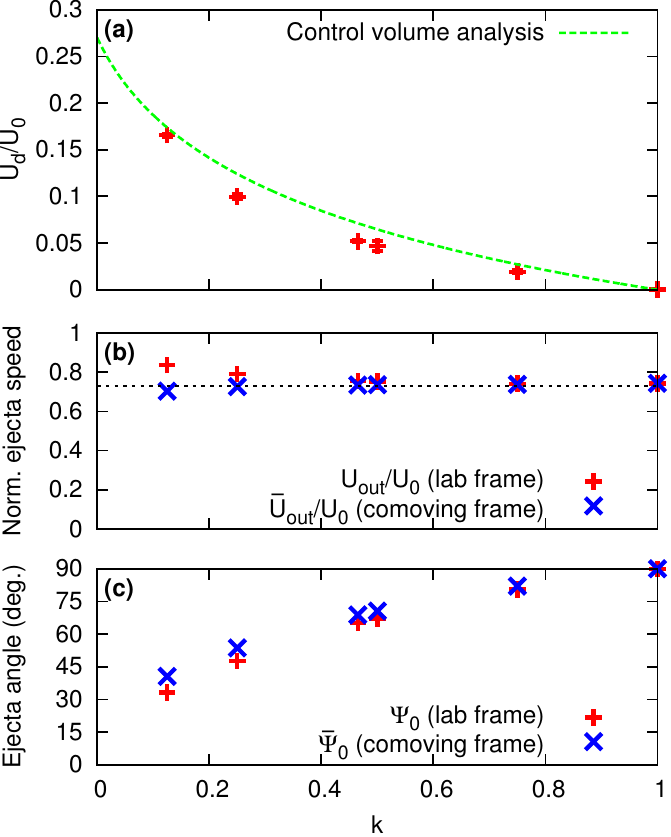}
\caption{\label{fig:udk}
(Color online) As the jets approach equal widths, the drift speed continuously vanishes. {\bf (a)} In the lab frame, we show the saddle point drift speed $U_d/U_0$ versus the ratio of the incident jet widths $k$ while keeping $R = 100R_{\rm G}$ and the restitution and friction coefficients $0.9$ and $0.2$, respectively in the discrete particle simulation (red points). We furthermore include the outcome from our control volume analysis for comparison, where we see the analysis perform quite well despite its simplicity. We see that $U_d$ decreases monotonically as $k$ increases, and eventually $U_d$ vanishes when the jets are equal widths at $k=1$. {\bf (b)} Ejecta velocity in the lab (red) and comoving (blue) frames. Note that the ejecta velocity appears relatively constant for all $k$, especially in the drifting frame. For the grains used here, that constant is approximately $B = 0.73$. {\bf (c)} To complete the characterization of the flow configuration resulting from the collision we include the ejecta angle in both the lab and comoving frames.
}
\end{figure}

For the discrete particle jets with coefficients of restitution and friction of $0.9$ and $0.2$, we varied the jet-width-ratio $k$ and observed the resulting impact configuration in the lab frame to provide a basic understanding of head-on jet impact response. This also allows us to verify that the simulations are producing sensible outcomes, for example, to verify that there is no central drift when the incident jets have equal widths.

The impact configuration is characterized by the drift speed $U_d$, the ejecta speed $U_{\rm out}$, and the ejecta angle $\Psi_0$. As later shown, the ejecta width $R_{\rm out}$ is determined uniquely from these parameters by mass conservation (Eq.~\eqref{eq:mass_flux_drift}).

In Fig.~\ref{fig:udk}(a) we see how the drift speed $U_d$ varies as $k$ is varied. When the jets are the same width, there is no drift, as one would expect due to symmetry. When the small jet is $0.465$ times the width of the large jet, as mentioned earlier, there is a drift speed of $U_d = 0.052U_0$, and when the small jet is $0.125$ times the width of the large jet, the drift speed increases to $U_d = 0.17 U_0$. For any $k < 1$ we found the drift emerge; there is not a threshold or yield $k$ above which there is no drift and below which it emerges. Rather the drift speed as a function of $k$ appears simply as an emergent result of unequal jet widths.

In Fig.~\ref{fig:udk}(b) we show the dependence of the ejecta speed $U_{\rm out}$ on $k$. Here we see much less variation than we did for the drift speed. Evidently, at fixed jet parameters, the ejecta speed does not change much even though the jet sizes are changed dramatically. In the cases presented, the smaller jet varied in width by a factor of $8$ while $U_{\rm out}$ varies by just $13\%$ in the lab frame.

Furthermore, we provide the ejecta angle $\Psi_0$ as the final ingredient to characterize the impact flow configuration. We see that $\Psi_0$ decreases as $k$ decreases, indicating that when the larger jet is much wider than the smaller one, the ejecta are more acute and are less deflected than the directly vertical jets obtained at the maximum deflection of $90^{\circ}$ when $k=1$.

\subsection{Bulk momentum absorption}

\begin{figure}[ht]
\centering
\includegraphics[width=\columnwidth]{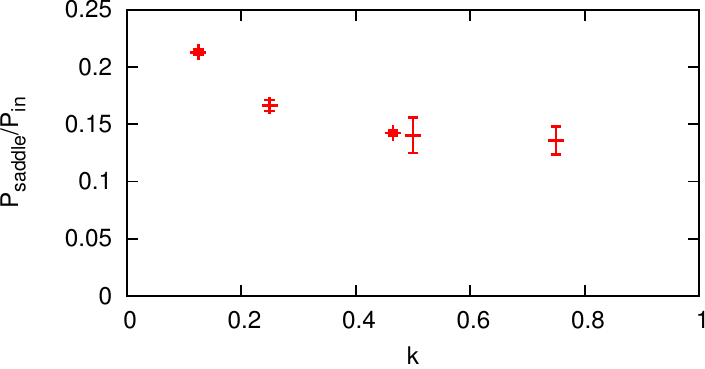}
\caption{\label{fig:momentum}
(Color online) Momentum balance incorporates momentum absorption into the drifting saddle. Here we show the momentum absorbed by the drifting saddle $P_{\rm saddle}$ as a fraction of the net incident momentum $P_{\rm in}$ as we vary the jet-width-ratio $k$. We see that the drifting saddle absorbs a considerable fraction of the net incident momentum. In the discrete particle jets used here, at least $15\%$ of the net incident momentum is captured by the drifting saddle.
}
\end{figure}

We now show that the drifting mode absorbs a significant fraction of net incident momentum. This absorption corresponds to a new term in the momentum balance for head-on jet collisions. Let us begin by defining the momentum fluxes in the three parts of the collision: $P_{\rm in}$ is the incident momentum flux, $P_{\rm out}$ is the momentum flux of the ejecta, and $P_{\rm saddle}$ is their difference:
    \begin{equation}
    P_{\rm in} - P_{\rm out} = P_{\rm saddle}
    \end{equation}
Importantly, because of the $y$-symmetry of our impact system, $P$ in all cases above refers only to the horizontal momentum along the $x$-axis. 

In the absence of drift, such as in ideal fluid impact, we know that the ejecta balance the momentum equation and $P_{\rm out} = P_{\rm in}$. However, in the case of granular jets, or drifting jets in general, we observe $P_{\rm saddle}$ entering the picture: by measuring the net incident momentum $P_{\rm in}$ and the net momentum in the ejecta $P_{\rm out}$ in our simulations, we find that their difference is nonzero. Namely, there is momentum being deposited into the saddle $P_{\rm saddle}$.

In Fig.~\ref{fig:momentum} we show the fraction of the initial momentum that is absorbed by the saddle $P_{\rm saddle}/P_{\rm in}$ as a function of the jet width ratio $k$. We see that indeed a significant fraction of the new incident momentum is being absorbed by the saddle, in this case roughly $15\%$ for the range of $k$ presented. 

This absorption is due to the fact that the flow field is uniformly translating, which in the lab frame appears simply as an elongation of the larger jet and a contraction of the smaller jet. Hence, in a unit time, the change in momentum due to the large jet elongation is $2\rho R U_0 U_d$, and the change due to the contraction of the smaller jet is $2 \rho k R U_0 U_d$. Thus
    \begin{equation}
    P_{\rm saddle} = 2 \rho R U_0 U_d (1 + k)
    \end{equation}
This effect is not unique to the momentum. In fact, the mass and energy also display this behavior, for example, the same approach yields a new term in mass balance due to the saddle motion given by $M_{\rm in} - M_{\rm out} = M_{\rm saddle} = 2 \rho R U_d (1 - k)$.

\subsection{Drift dependence on energy dissipation}

\begin{figure*}
\centering
\includegraphics[width=\textwidth]{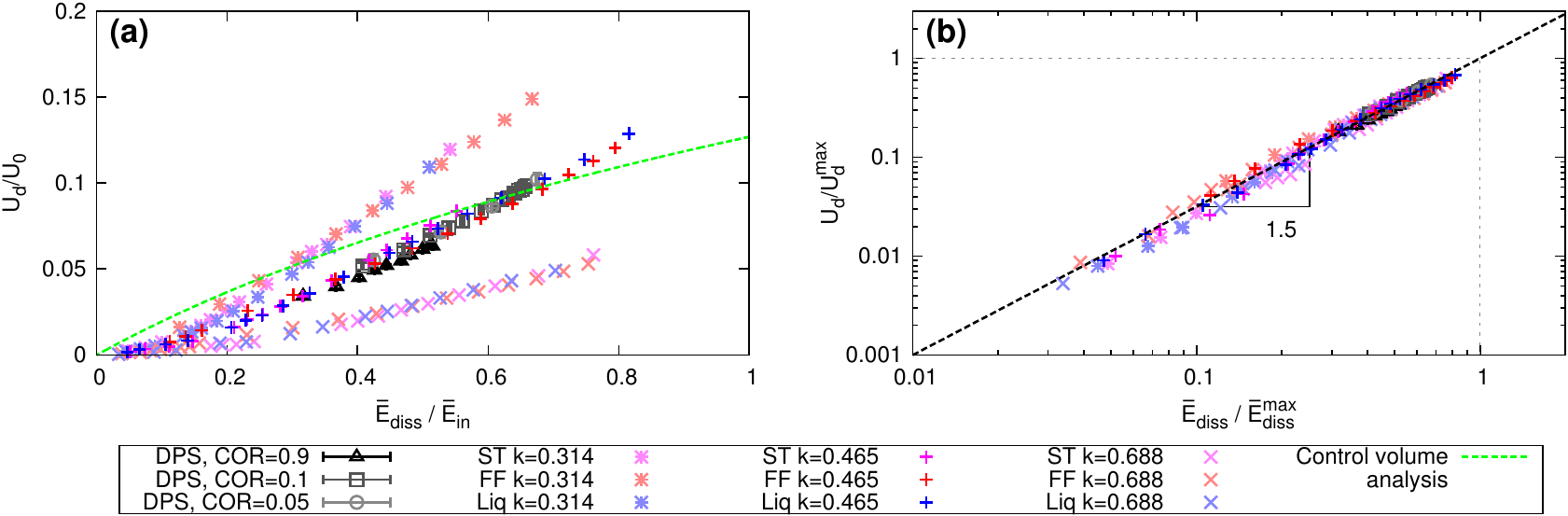}
\caption{\label{fig:ud_diss}
(Color online) Drift speed reaches a fraction of its maximum based only on how much of the available incident energy is dissipated. This fraction is achieved entirely independently of all jet-width-ratios and dissipation mechanisms. Shown in {\bf (a)} is the drift speed $U_d/U_0$ versus the normalized energy dissipation (from Eqs.~\eqref{eq:energy_dissipation} and \eqref{eq:in_energy}) for the discrete particle simulations and the continuum fluids. 
The discrete particle simulations (DPS) with coefficient of restitution (COR) 0.9 (black, triangles), 0.1 (dark grey, squares), and 0.05 (light gray, circles) are shown for $k = 0.465$. 
The continuum impacts are shown with $k = 0.314$ (\rlap{+}{\texttimes}), $k=0.465$ (+), and $k = 0.688$ (\texttimes) for the shear thickening fluid (ST, magenta), the frictional fluid (FF, red), and the Newtonian (Liq, blue) cases. At any given $k$, all of these cases produce similar drift speeds at a given energy dissipation rate. 
For comparison, we additionally include our control volume analysis (green, dashed) at $k = 0.465$, which qualitatively captures the increasing drift speed with increasing dissipation, but is completely unable to capture other features, most notably the concavity. {\bf (b)} Drift speed $U_d$ normalized by the maximum drift $\bar U_d^{\rm max}$ versus the energy dissipation rate $\bar E_{\rm diss}$ normalized by its maximum $\bar E_{\rm diss}^{\rm max}$. Firstly, we find a dramatic collapse of the data, indicating that when the drift speed reaches a given fraction of its maximum dependent only on the dissipation rate and independent of the details of the highly varying impact materials and jet-widths-ratios. Secondly, we find that this fraction is robustly dependent on the normalized dissipation rate to the power $1.5$.
}
\end{figure*}

Now we show that the main factor controlling the central drift is energy dissipation. For many jet-width-ratios, we vary the energy dissipation in the granular jet by changing the coefficient of restitution and coefficient of friction of the grains. We furthermore are able to explore the drift by varying dissipation in the impact of continuum fluids using the {\sc gerris} flow solver, where frictional-fluid, Newtonian and shear thickening constitutive relations are considered. 

The energy dissipation rate is most easily computed in the comoving reference frame. As mentioned, there are net mass, momentum, and energy transfers to the drifting center in quasi-steady-state found in the lab frame, and for simplicity we do our analysis in the comoving frame where a steady state is reached to simplify the analysis. In the comoving reference frame, we denote $\bar U_{\rm B}$ and $\bar U_{\rm S}$, as the incident speeds of the large and small jets respectively, and $\bar U_{\rm out}$ and $\bar\Psi_0$ as the ejecta speed and ejecta angle, respectively. These are related to their values in the lab frame via 
    \begin{align}
    &\bar U_{\rm B} = U_0 - U_d \label{eq:ub}\\
    &\bar U_{\rm S} = U_0 + U_d \label{eq:us}\\
    &\bar U_{\rm out} = \sqrt{(U_{\rm out}\cos\Psi_0 - U_d)^2 + U_{\rm out}^2\sin^2\Psi_0} \\
    &\bar \Psi_0 = \tan^{-1}\frac{U_{\rm out}\sin\Psi_0}{U_{\rm out}\cos\Psi_0 - U_d}
    \end{align}
The jet cross sections $R_{\rm out}$, $R$, and $kR$ are the same in the comoving and lab reference frames. The transformations of the ejecta speed and angle in the lab versus the comoving frames can be seen in for representative impacts in Figs.~\ref{fig:udk}(b) and (c).

In the comoving reference frame, when the impact produces a steady-state flow configuration, the dissipated energy $\bar E_{\rm diss}$ is equal to the energy advected in $\bar E_{\rm in}$ sans the energy advected out $\bar E_{\rm out}$. These energy fluxes are given 
    \begin{equation}
    \label{eq:diss_simple_drift}
    \bar E_{\rm diss} = \bar E_{\rm in} - \bar E_{\rm out}
    \end{equation}
Beginning with the energy influx, $\bar E_{\rm in}$ is simply the sum of the energy input from both the large and small jets
    \begin{equation}
    \label{eq:in_energy}
    \bar E_{\rm in} = \rho R (\bar U_{\rm B}^3 + k \bar U_{\rm S}^3)
    \end{equation}
Next, the energy advected out by the two ejecta jets is
    \begin{equation}
    \bar E_{\rm out} = 2 \rho R_{\rm out} U_{\rm out}^3
    \end{equation}
We can simplify the above slightly by noting that there is no net mass flux, and that all of the mass entering is consequently leaving. Denoting the mass fluxes by $\bar M$, we therefore have $\bar M_{\rm in} = \bar M_{\rm out}$, giving
    \begin{equation}
    \label{eq:mass_flux_drift}
    R (\bar U_{\rm B} + k \bar U_{\rm s}) = 2 \bar U_{\rm out} R_{\rm out}
    \end{equation}
because $\bar M_{\rm in} = 2 R (\bar U_{\rm B} + k \bar U_{\rm A})$ and $\bar M_{\rm out} = 4 \rho R_{\rm out} \bar U_{\rm out}$. Throughout the above we noted that the density $\rho$ of the jets remains constant during the impact in dense granular jet collisions~\cite{oblique}.

Combining Eqs.~\eqref{eq:diss_simple_drift} and \eqref{eq:mass_flux_drift}, we obtain
    \begin{equation}
    \label{eq:energy_dissipation}
    \bar E_{\rm diss} = \rho R\left[\bar U_{\rm B}^3 + k\bar U_{\rm S}^3-(\bar U_{\rm B} + k \bar U_{\rm S})\bar U_{\rm out}^2\right]
    \end{equation}

In Fig.~\ref{fig:ud_diss} we show how $U_d$ varies based on the fraction of energy dissipated during impact by comparing the dissipated energy \eqref{eq:energy_dissipation} to the input power $\bar E_{\rm in}$. This dependence is displayed for 4 different cases: the discrete particle granular jets like those in our previous results, and the three continuum fluids computed in {\sc gerris}. The continuum fluids we used were a frictional fluid model of dense granular jets~\cite{still_water, oblique}, a continuum Newtonian impact, and a continuum shear thickening fluid. The continuum impacts were repeated for $k = 0.314$, $k=0.465$ and $k=0.688$ while the discrete case is only presented for $k = 0.465$.

We varied the dissipation rate in the discrete particle granular jets by specifying the coefficient of restitution at $0.05$, $0.1$, and $0.9$, and for each coefficient of restitution, we varied the friction coefficient from $0$ to $1$. We varied the dissipation rate in the frictional fluid model by changing the dynamic friction coefficient $\mu$ from $0.05$ to $0.7$. Furthermore, we varied the dissipation rate in the Newtonian impact by varying the Reynolds number (${\rm Re}$) from $3$ to $100$ where the Reynolds number is defined ${\rm Re} = \rho R U_0/\eta$. Lastly, dissipation was varied in the shear thickening fluid by varying the effective Reynolds number ${\rm Re}_{\rm eff} = \rho R^{3/2} U_0^{1/2}/\kappa$  from $6$ to $100$. For the frictional fluid and Newtonian continuum simulations, we verified that the dissipation rate showed reasonable behavior as $\mu$ and $\eta$ were varied~\cite{supplemental}. 

In Fig.~\ref{fig:ud_diss}(a) we see that increasing the energy dissipation rate relative to the input power in the comoving frame, or increasing $\bar E_{\rm diss}/\bar E_{\rm in}$, increases $U_d$. Also, in the asymptotic case of no dissipation, the data shows that the drift speed itself also vanishes entirely. Hence, increasing dissipation increases the drift speed, and having no dissipation, such as in the head-on impact of ideal fluids, there is no drift. Also, in Fig.~\ref{fig:ud_diss}(a) we see that, consistent with Fig.~\ref{fig:udk}(a), reducing $k$ increases $U_{\rm d}$.

For all of these fluids, we found that the dependence of $U_d$ on $\bar E_{\rm diss}/\bar E_{\rm in}$ all coincide and collapse together at a given jet-width-ratio. The drift speed resulting from the head-on impact of unequal-width jets is insensitive to the precise {\em type} of the dissipation, rather, it only depends on the {\em amount} of dissipation in the jets for a given $k$. The mechanisms by which a Newtonian fluid dissipates energy are different from the generalized Newtonian shear thickening fluid, and in turn entirely different from that in a discrete particle granular jet. It is remarkable how the system produces such similar drift speeds for such different jet compositions hen controlled for the amount of dissipation. 

\subsection{Maximum drift speed and collapse}

Here we briefly derive the maximum drift speed available to the impact. Using the maximum drift speed as a characteristic velocity scale, we will then show that the scaled drift speed versus the scaled dissipation rate collapses across an enormous range of simulations. 

The drift speed $U_d$ is clearly a strictly increasing function in the dissipation rate, and so it is clear that the maximum drift speed is obtained when the dissipation is maximized. Here, the maximum energy dissipated $\bar E_{\rm diss}^{\rm max}$ is simply bounded by the energy supplied to the system:
    \begin{equation}
    \label{eq:diss_max}
    \bar E_{\rm diss}^{\rm max} = \bar E_{\rm in}      
    \end{equation}
Next, note that when drift is present, the momentum differential of the incident streams is lower in the comoving frame than in the lab frame. In the lab frame, the net incident momentum is 
    \begin{equation}
    P_{\rm in} = 2 \rho R U_0^2 ( 1 - k )
    \end{equation}
whereas in the comoving frame the net incident momentum is 
    \begin{equation}
    \bar P_{\rm in} = 2 \rho R ((U_0 - U_d)^2 - k (U_0 + U_d)^2) < P_{\rm in}
    \end{equation}
The question at hand is how far can the dissipation push the comoving incident momentum differential; the answer is until the two jets contain equal momenta in the comoving frame. This corresponds to directly vertical ejecta streams with $\bar \Psi_0 = \pi/2$. 
At the maximum energy dissipation, where the drift speed is maximized, $\bar P_{\rm in} = 0$ and 
    \begin{equation}
    (U_0 - U_d^{\rm max})^2 = k (U_0 + U_d^{\rm max})^2
    \end{equation}
yielding a maximum drift speed
    \begin{equation}
    \label{eq:ud_max}
    \frac{U_d^{\rm max}}{U_0} = \frac{1 - \sqrt{k}}{1 + \sqrt{k}}
    \end{equation}

Using $\bar E_{\rm diss}^{\rm max}$ as the characteristic energy scale and the corresponding $U_d^{\rm max}$ as the characteristic velocity scale, we see in Fig.~\ref{fig:ud_diss}(b) that our results do not only collapse across dissipation mechanisms for a given $k$, but further that all of our head-on impacts appear to be controlled by these characteristic velocity and energy scales. Hence when the dissipation is at a given fraction of the available energy $\bar E_{\rm diss}/\bar E_{\rm diss}^{\rm max}$, the drift speed is at a given fraction of its maximum $U_d/U_d^{\rm max}$ as well, independent of the jet-width-ratio $k$, and independent of the details of the dissipation mechanisms of the impact. Essentially, the emergent drift speed in the end appears to be a relatively simple phenomenon. The effects of the jet-width-ratio are easily captured based on the total available energy to the system to the extent where the drift speed is apparently determined by a self-similar equation
    \begin{equation}
    \label{eq:similarity}
    U_d = U_d^{\rm max} f (\bar E_{\rm diss}/\bar E_{\rm diss}^{\rm max})
    \end{equation}
where $f (0) = 0$, $f (1) = 1$, and $f$ is strictly increasing. 

\subsection{Drift speed scaling relative to the dissipation}

In Fig.~\ref{fig:ud_diss}(b), aside from the dramatic data collapse, we see that the drift speed has a power law scaling relative to the dissipation rate where $U_d \sim \bar E_{\rm diss}^{1.5}$. Though this scaling is clearly and apparently present, we do not have an explanation for the scaling at this time. In particular, it appears surprisingly that $U_d = U_d^{\rm max}(E_{\rm diss}/E_{\rm diss}^{\rm max})^{1.5}$ in Eq.~\eqref{eq:similarity}.

A particularly confounding aspect of the power law are that straightforward analyses predict nothing like an exponent of $1.5$. For example, in order to recover such a law using dimensional analysis, our analysis would have to yield, in some form
    \begin{equation}
    U_d^2 \sim E_{\rm diss}^3
    \end{equation}
However, a physically relevant dimensional quantity dependent on the cube of the dissipation rate is highly obscure, and it is unclear to us where one might arise. 

Furthermore, linear perturbations to the system, for example to the similarity equation Eq.~\eqref{eq:similarity} to infer the form for $f$, would not yield a $1.5$ or $3/2$ power, but rather a linear dependence of $f$ on the dissipation. For all of these reasons, the scaling identified in Fig.~\ref{fig:ud_diss} is puzzling.

Below we attempt to understand the results exposed above by presenting a very simple control volume analysis. As we will see, the analysis captures the broad nature of the effects of dissipation and unequal jet widths on the drift speed from our simulation, but is far from capturing the simple form for $f$ as we empirically identified from our simulations.

\section{Control Volume Analysis}

Here we provide a simple approximation of the drift speed resulting from the head on in pact of two unequal-width, dissipative jets as a qualitative exercise in showing how dissipation and an unequal jet-size-ratio can naturally produce a drift speed. This approximation is largely independent from measurements or observations from our simulations, providing a from-scratch look at the effects of dissipation and the jet-width-ratio.

We conduct our analysis in the drifting frame. The control volume analysis consists of two rough approximations: one regarding the ejecta speed, and the other approximating the energy dissipation rate. These two parts are related but distinct, since the ejecta speed depends not only on the dissipation rate, but also the drift speed.

\subsection{Approximating dissipation rate and ejecta speed}

In the following analysis, we presume that most of the dissipation occurs in the vicinity of the smaller jet near the impact saddle. This corresponds to a length scale $L \sim CkR$, a velocity scale $U\sim \bar U_{\rm S}$, and, in two dimensions, a volume scale of $V\sim L^2$. Here, the constant $C$ is assumed to be independent of the jet width ratio and the dissipation rate.

The control volume analysis begins with a very rough approximation of the energy dissipated during impact. The total energy dissipation rate, shown in Eq.~\eqref{eq:energy_dissipation}, relates the energy dissipation to the change in the incident and outgoing energy fluxes. More plainly, 
    \begin{equation}
    \bar E_{\rm diss} = -\oint \frac{\rho \bar u^2}{2} \bar{\v u}\cdot\d\v S
    \end{equation}
where the contour is along a control volume surrounding the jet and extending far from the impact center in order to enclose the entire flowing region with nonzero velocity gradients. This control volume is fixed in space since the flow in the comoving frame is in steady-state. This becomes, noting fluid incompressibility,
    \begin{multline}
    \label{eq:vol_dissipation}
    \oint \frac{\rho \bar u^2}{2} \bar{\v u}\cdot\d\v S =\\ \int \boldsymbol\nabla\cdot \left(\frac{\rho \bar u^2}{2}\bar{\v u}\right) \d V =
    \frac{\rho}{2} \int \bar {\v u} \cdot \boldsymbol\nabla (\bar u^2) \d V
    \end{multline}

Using dimensional analysis we can approximate $\bar E_{\rm diss}$ from Eq.~\eqref{eq:vol_dissipation} as
    \begin{equation}
    \label{eq:diss_dimensional_analysis}
    \rho U \frac{\Delta(\bar u)^2}{L} V \sim \rho C k R (\bar U_{\rm out}^2 - \bar U_{\rm S}^2)\bar U_{\rm S}
    \end{equation}
Above we considered an additional approximation where $\Delta(\bar u^2) = \bar U_{\rm out}^2 - \bar U_{\rm S}^2$, as this is the change in the velocity squared from our final velocity state relative to our dominant velocity scale. This, together with Eq.~\eqref{eq:energy_dissipation}, yields
    \begin{equation}
    \label{eq:model_diss_approx}
    \bar U_{\rm B}^3 + k\bar U_{\rm S}^3-(\bar U_{\rm B} + k \bar U_{\rm S})\bar U_{\rm out}^2
     = C k (\bar U_{\rm S}^2 - \bar U_{\rm out}^2)\bar U_{\rm S}
    \end{equation}

For equal-width jets when $k = 1$, symmetry requires $U_d = 0$, and hence $\bar U_{\rm S} = \bar U_{\rm B}$  $U_{\rm d} = 0$. In this case, Eq.~\eqref{eq:model_diss_approx} reduces to 
    \begin{equation}
    2 \bar U_{\rm S}^3 - 2 \bar U_{\rm S} U_{\rm out}^2 = C \bar U_{\rm S}(\bar U_{\rm S}^2 - \bar U_{\rm out}^2)
    \end{equation}
which requires
    \begin{equation}
    C = 2
    \end{equation}

The final part of the approximation concerns the ejecta speed. Our analysis simply approximates the ejecta speed in the comoving frame as 
    \begin{equation}
    \label{eq:b}
    \bar U_{\rm out} = B U_0
    \end{equation}
In the above approximation, we allow $B$ to vary when fluid properties vary, such as changing the Reynolds number in a Newtonian impact, or the friction coefficient in a discrete-particle granular impact. But, we presume that $B$ does not change as $k$ is varied. This approximation is motivated by Fig.~\ref{fig:udk}(b), where we see that $\bar U_{\rm out}$ is approximately constant for the range of $k$ tested. In that example with discrete particle jets whose coefficients of restitution and friction are respectively $0.9$ and $0.2$, we obtain $B = 0.73$.

Hence, with $C$ determined by symmetry, the analysis becomes dependent on the parameter $B$ above which accounts for dissipation effects. Combining Eqs.~\eqref{eq:model_diss_approx} and \eqref{eq:b}, our analysis simply yields
    \begin{equation}
    \label{eq:model}
    \bar U_{\rm B}^3 + k\bar U_{\rm S}^3-(\bar U_{\rm B} + k \bar U_{\rm S})B^2 U_0^2
     = 2 k (\bar U_{\rm S}^2 - B^2 U_0^2)\bar U_{\rm S}
    \end{equation}
The result of the control volume analysis, shown in Eq.~\eqref{eq:model} is a cubic equation for the drift speed. Below we explicitly show the cubic equation for $U_d$ from expanding Eq.~\eqref{eq:model} and recalling Eqs.~\eqref{eq:ub} and \eqref{eq:us}:
    \begin{multline}
    \label{eq:ud_cubic}
    \left(\frac{U_d}{U_0}\right)^3 -
    3\left(\frac{U_d}{U_0}\right)^2 \frac{1-k}{1+k} \\ +
    \frac{U_d}{U_0}(3 - B^2) -
    (1-B^2)\frac{1-k}{1+k} = 0
    \end{multline}
Because Eq.~\eqref{eq:ud_cubic} has a negative discriminant for virtually all $k$ and $B$, it in all practicality admits a unique real solution for $U_d$, and the drift speed can be determined relative to $U_0$ if provided $k$ and $B$.

\subsection{Analogy to equal dissipation in both incident jets}

Here we show that the energy dissipation approximation in Eq.~\eqref{eq:model_diss_approx} is equivalent to the approximation that an equal amount of energy is dissipated in fluid originating in the two incident jets. That is, our energy dissipation approximation in Eq.~\ref{eq:model_diss_approx} analogous to approximating the energy drop across fluid originating in the large incident stream as equal to that in the smaller incident stream.

Beginning with the small jet in the comoving frame, we know that the incident mass flux for particles originating in the small jet equals the outgoing mass flux for those same particles. Because the ejecta is at a speed $\bar U_{\rm out}$
    \begin{equation}
    \rho k R \bar U_{\rm S} = \rho R_{\rm S,out} \bar U_{\rm out}
    \end{equation}
where $R_{\rm S,out}$ is the width of the ejecta stream including fluid originating from only the small incident jet. As before, we calculate the energy dissipated in the small jet, denoted $\bar E_{\rm S,diss}$, via
    \begin{multline}
    \label{eq:eq_power_diss_small}
    \bar E_{\rm S,diss} = \rho kR \bar U_{\rm S}^3 - \rho R_{\rm S,out} \bar U_{\rm out}^3 \\= \rho kR\bar U_{\rm S} (\bar U_{\rm S}^2 - \bar U_{\rm out}^2)
    \end{multline}
Thus, if we set the energy dissipated in fluid from the large jet equal to that from the smaller jet $\bar E_{\rm B,diss} = \bar E_{\rm S,diss}$, we find that this approximation yields
    \begin{equation}
    \label{eq:sum_diss_big_small}
    \bar E_{\rm diss} = \bar E_{\rm S,diss} + \bar E_{\rm B,diss} = 2\bar E_{\rm S,diss}
    \end{equation}
Thus Eqs.~\eqref{eq:eq_power_diss_small} and \eqref{eq:sum_diss_big_small} recover Eq.~\eqref{eq:model_diss_approx}. Therefore approximating the power dissipated in both jets as being equal is equivalent to the rough dimensional approximation done earlier.

\subsection{Evaluation of the control volume analysis}

We now discuss how drift emerges amid dissipation in our control volume analysis, as well as compare the analysis to our discrete particle and continuum jet impact simulations. In the comparisons against previous results, we first compare the control volume analysis to our discrete particle simulations with fixed particle properties (or fixed $B$) and varied $k$. We then test how the analysis captures the effect of dissipation by fixing $k$ and varying the dissipation by changing $B$. 

We first identify qualitative outcomes pertaining from the head-on collisions of two streams in the context of the control volume analysis from Eq.~\eqref{eq:model}. Namely, we naturally find the presence of dissipation produces a drift speed. In the absence of dissipation, the jets behave as ideal fluids, and thus $B = 1$ because the ejecta speed must equal the incident speed. In that case, Eq.~\eqref{eq:model} reduces to 
    \begin{equation}
    \bar U_{\rm B} (\bar U_{\rm B} - U_0^2) = k \bar U_{\rm S} (\bar U_{\rm S} - U_0^2)
    \end{equation}
whose only real solution is obtained when $\bar U_{\rm B} = \bar U_{\rm S} = U_0$, and no drift is present. Decreasing $B$ from unity corresponds to increasing dissipation, where $\bar U_{\rm B} < \bar U_{\rm S}$, and a positive drift speed emerges.

Second, we check how the control volume analysis performs when fixing $B$ and varying $k$. In particular, we check how the analysis performs in our granular jet simulations with coefficients of restitution and friction of $0.9$ and $0.2$ at various $k$. In Fig.~\ref{fig:udk}(a) we show the drift speed versus $k$ for the granular jets as well as for the control volume analysis with $B = 0.73$, as obtained from the granular jet noting that in the comoving frame, $\bar U_{\rm out} = 0.73 U_0$ (Fig.~\ref{fig:udk}(b)). We see that the analysis captures the trend in $U_d$ when $B$ is fixed and $k$ is varied.

The final check for our control volume analysis is when the jet-width-ratio is fixed at $k = 0.465$ and the dissipation is varied by changing $B$ from $0$ to $1$. 
The comparison of the analysis to our various simulations is found in Fig.~\ref{fig:ud_diss}(a). The analysis captures the broadest qualitative nature found in the simulations, namely that there is no drift speed in the absence of dissipation and that increasing the dissipation rate monotonically increases the drift speed. However, the analysis fails to capture the clear concavity found in the multiple collapsing simulation data sets. As such, the analysis fails to capture the $1.5$ exponent in the scaling identified in Fig.~\ref{fig:ud_diss}(b). 

To its credit, the analysis does not have much tunability, and it nevertheless is in order the simulation results. 
However, we find that the analysis fails quantitatively. 
Its largest shortcoming is when comparing the drift speed versus the dissipation rate, where the analysis predicts the wrong concavity for $U_d$ versus $\bar E_{\rm diss}$. and therefore does not capture the robust scaling that we identified in our simulations.

\section{Conclusion}

The impact of ideal fluid jets in two dimensions is a classic pedagogical problem in fluid dynamics, both because it is a conceptually simple generalization of the classical mechanics particle collision problem to fluid dynamics and because it is analytically tractable using classical conformal mapping techniques~\cite{birkhoff, gurevich, milne-thomson}. While it has long been appreciated that flow reversibility in ideal, dissipation-free jet impact dictates that the central impact region remains fixed in the laboratory frame, there has been no study focused on how dissipation affects this outcome.

In this study, we investigated the effects of dissipation in two dimensional head-on jet impacts, focusing on the head-on impact of granular jets. We identified that the collision produces a steadily drifting flow configuration which absorbs a significant fraction of the incident momentum into the collectively drifting impact center. We further showed that energy dissipation is the main ingredient necessary to produce the emergent drift. To this point, we saw that the amount of dissipation dictates what fraction of the maximum drift speed is reached across a very wide variety of fluid dissipation mechanisms and jet-width-ratios, and identified that the drift speed robustly scales relative to the energy dissipation rate with a $1.5$ exponent. We lastly motivated a simple control volume analysis of the collision which captured the general effect of dissipation on the drift speed, though this analysis was short of explaining the scaling identified in our simulations.

The findings presented above, together with ejecta insensitivity to internal structure in jet-target impacts begin to paint a picture of extreme ejecta insensitivity to fluid composition during dense jet impacts. In the case considered in this study, multiple different types of fluids, from discrete particle to shear thickening, all produce the same drift speeds from head-on collisions when controlled for total dissipation; the precise form of the dissipation evidently does not affect the outcome of the impact, rather the only determining factor is the amount of dissipation. Corroborating this insensitivity in dense jet impact in~\cite{still_water}, when impinging granular and ideal fluid jets against a fixed target, the internal structures showed very weak signatures in the ejecta. Thus, it appears that the ejecta resulting from impact of dense jets holds a very weak signature regarding the internal structure or composition of the impact. This challenges the traditional scattering approach in which the ejecta are used to infer material properties or internal structure, such as has been done to identify the double-helix structure of DNA, the subatomic structure of Baryonic Matter in high energy physics, as two of several examples.

Furthermore, when investigating the impact between two free streams or objects, it is natural to simplify the dynamics by fixing the first object and then collide the second against the first. Examples where this has has been done when studying the off-center collision of granular streams~\cite{xiang_oblique} or the formation of protoplanets~\cite{teiser_wurm_main, blum_wurm_review, wurm_paraskov_krauss_early}. However, because of the presence of central drift in dense, dissipative impacts, the outcome of collisions when fixing one object is fundamentally different than the free-collision counterpart. This consideration can impact the interpretation, and perhaps even the validity, of some fixed-target impact experiments mimicking the high-speed free collision of two dense objects. For example, since the decimation of a dust aggregate is an important consideration in protoplanetary formation, the total force felt during impact is a crucial experimental output. But, in this case, fixing one object produces larger impact forces than free impacts because a free impact would reduce the collision force via the central drift.

We would also like to highlight that the scaling behavior of the drift speed relative to the energy dissipation rate remains a mystery. Our simple approaches were not able to capture the exponent, and the similarity equation or control volume analysis presented cannot capture the scaling short of significant insight. In the control volume analysis, we assumed that the dissipation occurred over a length scale dependent only on $k$ and not the energy dissipated. But, evidently the dissipation length scale is more sophisticated, and our evidence suggests that the relevant impact length scale varies with the dissipation rate. This, taken together with the fact that the details of the jet constitutive material are irrelevant, lay the groundwork for a more involved analysis with a more careful accounting of the impact length scale. In particular, this would manifest as a the coefficient $C$ in Eq.~\eqref{eq:model_diss_approx} as no longer being constant, but varying based on the dissipation.

Lastly, we remark that though we believe drift should occur in three-dimensional axisymmetric impacts, we are unsure if it too is insensitive to material type or results in a simple scaling behavior relative to the dissipation rate. The control volume analysis was clearly able to capture the qualitative nature of the central drift speed, and this analysis can easily be extended to show the same result in axisymmetric impacts. Whether or not the finding here that the total dissipation is significantly more important than the material composition is unclear, as the analysis was unable to entirely capture this effect, and a three-dimensional study with numerical diligence is required to determine if all of the two-dimensional findings in this paper persist.

\acknowledgments
The work was supported by the National Science Foundation through its Fluid Dynamics Program (CBET-1336489 to PI Wendy W. Zhang). The author would like to particularly thank W.~W.~Zhang for discussions, insights, and careful readings of this paper, in addition to T.~A.~Witten, H.~M.~Jaeger, S.~R.~Nagel, I.~Bischofberger, N.~Guttenberg and M.~Z.~Miskin for insights and discussions.

\bibliography{headon}

\end{document}